\documentclass[prd,aps,twocolumn,floatfix,superscriptaddress]{revtex4}

\def\rmd{{\rm d}}
\def\p{\partial}

\usepackage{amssymb,graphicx,color}

\begin{document}
%%%%%%%%%%%%%%%%%%%%%%%%%%%%%%%

\title{A remedy for constraint growth in Numerical Relativity}

\author{Gioel Calabrese}

\affiliation{School of Mathematics, University of Southampton,
Southampton, SO17 1BJ, UK}

\date{\today}

%%%%%%%%%%%%%%%%%%%%%%%%%%%%%%%%%%%%%%%%%%%%%%%%%%%%%%%%%%%%%
\begin{abstract}
%%%%%%%%%%%%%%%%%%%%%%%%%%%%%%%%%%%%%%%%%%%%%%%%%%%%%%%%%%%%%

Rapid growth of constraints is often observed in free evolutions of
highly gravitating systems.  To alleviate this problem we investigate
the effect of adding spatial derivatives of the constraints to the
right hand side of the evolution equations, and we look at how this
affects the character of the system and the treatment of boundaries.
We apply this technique to two formulations of Maxwell's equations,
the so-called fat Maxwell and the Knapp-Walker-Baumgarte systems, and
obtain mixed hyperbolic-parabolic problems in which high frequency
constraint violations are damped.  Constraint-preserving boundary
conditions amount to imposing Dirichlet boundary conditions on
constraint variables, which translate into Neumann-like boundary
conditions for the main variables.  The success of the numerical tests
presented in this work suggests that this remedy may bring benefits to
fully nonlinear simulations of General Relativity.

\end{abstract}

\maketitle

%\tableofcontents

%%%%%%%%%%%%%%%%%%%%%%%%%%%%%%%%%%%%%%%%%%%%%%%%%%%%%%%%%%%%%%
\section{Introduction}
%%%%%%%%%%%%%%%%%%%%%%%%%%%%%%%%%%%%%%%%%%%%%%%%%%%%%%%%%%%%%%

The $3+1$ decomposition of Einstein equations gives rise to a system
of evolution equations and a set of constraints. Although it is true
that, at the analytical level and in the absence of boundaries, if the
constraints are satisfied on an initial spacelike hypersurface, they
remain satisfied at later times by virtue of the evolution equations,
this is not the case in numerical simulations.  If the continuum
initial data satisfy the constraints exactly, then they will only
satisfy the discretized constraints up to truncation error.  In many
cases, initial errors grow exponentially, and sometimes even more
rapidly.  Whether the violent growth of the constraints is causing the
crash of numerical codes, or is simply the effect of a more general
growth in the error, is not altogether clear.  However, the
hope is that, if the constraint growth can be prevented or at
least alleviated, the runtime of numerical simulations may
increase.

In recent years a number of techniques have been proposed aimed at
damping the constraints or, at least, controlling their growth.  These
are the $\lambda$-system \cite{lambda}, the fine tuning of parameters
multiplying the constraints that are added to the evolution equations
\cite{SKLPT}, the dynamical adjustment of such parameters \cite{T},
and the addition of variational derivatives of a constraint energy to
the evolution equations \cite{F}.  The method that we wish to
investigate has remarkable similarities with the latter.  When
timelike boundaries are present all of these methods require
constraint-preserving boundary conditions to prevent the injection of
constraint violations.

We begin with a simple observation.  Consider the first order scalar  
equation on the real line,
\begin{equation}
u_t = i u_x\,,
\label{Eq:illposed}
\end{equation}
where $i^2 = -1$, supplemented with smooth initial data.  If we make
the ansatz $u(t,x) = e^{i\omega x}\hat{u}(t,\omega)$, $\omega \in
\mathbb{R}$, and substitute it into (\ref{Eq:illposed}), we obtain
$\hat{u}_t = -\omega \hat{u}$.  The fact that the solution, $u(t,x) =
e^{i\omega x -\omega t} \hat u(0,\omega)$, allows for unbounded growth
\footnote{By this we mean that it is not possible to find two
  constants $K>0$, $\alpha \ge 0$, such that the bound $\|u(t,\cdot)\|
  \le K e^{\alpha t} \| u(0,\cdot)\|$ holds for any smooth initial
  data.}  shows that the initial value problem is ill-posed and,
  therefore, cannot be consistently discretized in a stable way.  Now
  we modify our problem and assume that the constraint
\begin{equation}
C \equiv u_x +iu = 0 \label{Eq:illposed2}
\end{equation}
has to hold.  Clearly, if $u$ satisfies Eq.~(\ref{Eq:illposed}), then
$C_t = iC_x$, i.e., the constraint propagation problem is also
ill-posed.  A crucial point is that we are allowed to add constraints
and/or derivatives of constraints to the right hand side of the
evolution equations without affecting the space of physical solutions
(i.e., solutions that satisfy the constraints).  For example, adding a
spatial derivative of the constraint to the right hand side of
Eq.~(\ref{Eq:illposed}) leads to
\begin{equation}
u_t = i u_x + C_x = 2i u_x + u_{xx}\,. \label{Eq:illposed3}
\end{equation}
This equation is parabolic and, most importantly, gives rise to a
well-posed initial value problem for both the main and the constraint
propagation systems.  Not only does the problem no longer suffer from
unbounded growth of the constraints, but the modification also damps
the high frequency constraint violations.  Furthermore, in the
presence of boundaries, we can impose homogeneous Dirichlet boundary
conditions on the constraint, $C=0$, which, when translated in terms
of the main variables become of Neumann type, $u_x = -iu$. This
example, although trivial, clearly illustrates the potential benefits
of dissipation.  It allowed us to convert an ill-posed initial value
problem into a well-posed initial-boundary value problem \footnote{The
addition of the term $-iC$ to the right hand side of the evolution
equation (\ref{Eq:illposed}), would also have led to a well-posed
initial-boundary value problem.  However, this modification also gives
rise to exponential growth of all Fourier components of the
constraints.}.

In this paper we investigate the effect of adding spatial derivatives
of the constraints to the evolution equations for two formulations of
Maxwell's equations, the so-called fat Maxwell \cite{L} (first order)
and the Knapp-Walker-Baumgarte (KWB) \cite{KWB} (first order in time,
second in space) systems.  As expected, the modification destroys the
hyperbolic character of the systems.  However, if done carefully, it
can lead to mixed hyperbolic-parabolic systems both for the evolution
equations and the evolution of the constraints, such that the high
frequency components of the constraints are damped.

In the presence of timelike boundaries the systems that we analyze are
such that constraint-preserving boundary conditions amount to imposing
homogeneous Dirichlet boundary data on the constraints, which
translate into Neumann-like boundary conditions for the main system.
Compared to standard constraint-preserving boundary conditions for
hyperbolic problems \cite{IR,CLT,CPSTR,SW,Stewart}, those that we
obtain can be much simpler and possibly more robust.  For example, if
the constraint propagation system is strongly parabolic, then at the
boundary we can impose that all constraints vanish, which is less
involved than computing the incoming characteristic constraints and,
as usually done, trading the normal with time and tangential
derivatives using the evolution equations.  Further, the modified
problems allow for a greater number of constraint components to be set
to zero.  Unfortunately, --- and this is presumably the most
regrettable feature of this technique --- due to the loss of finite
speed of propagation, the time step needs to be proportional to the
square of the mesh spacing in order to achieve numerical stability
(and convergence) with explicit finite difference schemes.

Parabolic systems have received little attention in numerical
relativity with some notable exceptions \cite{parab}, in which the aim
was primarily to promote elliptic gauge conditions to parabolic
equations.  Although the technique of adding spatial derivatives of
the constraints has been previously employed by Yoneda and Shinkai
\cite{YS} and by Fiske \cite{F}, the analysis of the character and the
study of appropriate boundary conditions for the modified systems, to
our knowledge, is new.

We recall some basic definitions that will be used throughout the
paper.  Consider the systems of partial differential equations in one
space dimension
\begin{eqnarray}
\p_t u &=& A \p_x^2 u + B\p_x u + C u\,,\label{Eq:parabolic}\\
\p_t v &=& D\p_x v + E v\,, \label{Eq:hyperbolic}
\end{eqnarray}
where $A$, $B$, $C$, $D$, and $E$ are constant square matrices.
System (\ref{Eq:parabolic}) is said to be {\em strongly parabolic} if
the eigenvalues of $A+A^{\dagger}$ are positive.  We call system
(\ref{Eq:hyperbolic}) {\em symmetric hyperbolic} if $D = D^{\dagger}$.
In the presence of boundaries and with appropriate initial and
boundary conditions both systems admit a well-posed initial-boundary
value problem as does the coupled {\em mixed hyperbolic-parabolic}
system \cite{GKO}
\begin{equation}
\p_t \left(\begin{array}{c} u \\ v \end{array} \right) = 
\left( \begin{array}{cc} A \p_x^2 + B\p_x + C & F\p_x + G\\
L\p_x + M & D\p_x + E \end{array} \right)
 \left(\begin{array}{c} u \\ v \end{array} \right) \label{Eq:mixed}
\end{equation}
for any choice of the coupling matrices $F$, $G$, $L$, and $M$.
Notice that the character of system (\ref{Eq:mixed}) is entirely
determined by the two matrices $A$ and $D$.

The paper is organized as follows.  In Sec.~\ref{Sec:fat} we introduce
the fat Maxwell system, appropriately add spatial derivatives of
constraints and analyze the resulting initial-boundary value problem.
A similar analysis is then repeated for the KWB system in
Sec.~\ref{Sec:KWB}. In Sec.~\ref{Sec:numerics}, after describing the
discretization of the system and, particularly, of the boundary
conditions, we summarize the results of our numerical experiments.  We
conclude with some final remarks in Sec.~\ref{Sec:conclusion}.  In
appendix \ref{App:discretization} we discuss an energy conserving
discretization for the wave equation in second order form, allowing
for Sommerfeld boundary conditions.  This discretization is used for
the hyperbolic sector of the KWB system.

We use units where $c=\mu_0=\epsilon_0 = 1$ and adopt the Einstein
summation convention.

%%%%%%%%%%%%%%%%%%%%%%%%%%%%%%%%%%%%%%%%%%%%%%%%%%%%%%%%%%%%%%
\section{The fat Maxwell system}
\label{Sec:fat}
%%%%%%%%%%%%%%%%%%%%%%%%%%%%%%%%%%%%%%%%%%%%%%%%%%%%%%%%%%%%%%

The vacuum Maxwell equations on a Minkowski background in Cartesian
coordinates are given by
\begin{eqnarray}
\p_t E_i &=& \epsilon_{ijk} \p^jB^k\,, \label{Eq:max1}\\
\p_t B_i &=& -\epsilon_{ijk} \p^jE^k\,, \label{Eq:max2}\\
\p^kE_k &=& 0\,,\label{Eq:max3}\\
\p^kB_k &=& 0\,.\label{Eq:max4}
\end{eqnarray}
In terms of the potentials, $\phi$ and $A_i$, the electric and
magnetic fields are 
\begin{eqnarray*}
E_i &=& -\p_t A_i -\p_i\phi\,, \\
B_i &=& \epsilon_{ijk} \p^jA^k\,.
\end{eqnarray*}
If we introduce the variables $D_{ij} = \p_iA_j$, we can rewrite Maxwell's
equations in the form \cite{L}
\begin{eqnarray*}
\p_t E_i &=& \p^k (D_{ik} - D_{ki})\,,\\
\p_t D_{ij} &=& - \p_iE_j - \p_i\p_j \phi\,,\\
C & \equiv & \p^k E_k = 0 \,,\\
C_{ijk} & \equiv & \p_i D_{jk} - \p_j D_{ik} = 0\,,
\end{eqnarray*}
where we used the identity $\epsilon_{ijk}\epsilon^{ilm} =
\delta^l_j\delta^m_k - \delta^m_j \delta^l_k$.
This system has 12 variables that depend on the space-time coordinates
and is subject to 10 constraints.  One can readily verify that the
constraints propagate trivially,
\begin{eqnarray*}
\p_t C &=& 0\,,\\
\p_t C_{ijk} &=& 0\,,
\end{eqnarray*}
and that the evolution system is weakly hyperbolic, i.e., it cannot be
diagonalized.  However, by appropriately adding constraints to the
right hand side of the evolution equations
\begin{eqnarray*}
\p_t E_i &=& \p^k (D_{ik} - D_{ki}) + C_{ikk}\\
&=&- \p^k D_{ki}+ \p_i D_{kk}\,,\\
\p_t D_{ij} &=& - \p_iE_j -\p_i\p_j\phi + \delta_{ij} C\\
&=& - \p_iE_j -\p_i\p_j\phi + \delta_{ij} \p^kE_k\,,
\end{eqnarray*}
one can ensure that both the main evolution system and the evolution
of the constraint variables are symmetric hyperbolic.  We note that
this modification corresponds to setting $\gamma_1 = 2$ and $\gamma_2
= 1$ in the two parameter family of formulations of Lindblom et
al.~\cite{L}.

Any first order, linear, homogeneous, constant coefficient, symmetric
hyperbolic system with no lower order terms admits, by definition, a
conserved energy.  An energy, in this context, is an integral over the
spatial domain, $\Omega$, of a positive definite quadratic form of the
main variables.  The energies
\begin{eqnarray}
E &=& \frac{1}{2}\int_{\Omega} \left( E^iE_i +D^{ij}D_{ij} \right) \rmd^3 x \,,\label{Eq:E}\\
E_C &=& \frac{1}{2}\int_{\Omega} \left(C^2 + \frac{1}{2}C^{ijk}C_{ijk} \right)
\rmd^3 x\,, \label{Eq:EC}
\end{eqnarray}
for example, are conserved up to contributions due to the boundary
(the conservation of $E$ requires the Coulomb gauge, $\phi = 0$, which
will henceforth be assumed).  An immediate consequence of this is
that, if $\Omega = \mathbb{R}^3$ or $\Omega = T^3$, the energy
associated with an initial violation of the constraints will be
preserved during evolution.  We now show that by adding spatial
derivatives of the constraints to the evolution equations it is
possible to improve on this.

The general idea is to add constraints and/or derivatives of constraints
to the right hand side of the evolution equations that generate
decaying or damping terms in the constraint propagation system.
Consider the following modification of the main evolution system
\begin{eqnarray}
\p_t E_i &=& \ldots  + \mu \p_iC \label{Eq:Et}\\
&=& - \p^k D_{ki}+ \p_i D_{kk} + \mu \p_i\p^kE_k\,,\nonumber\\
\p_t D_{ij} &=& \ldots + \mu \p^k C_{kij}\label{Eq:Dt}\\
&=& - \p_iE_j + \delta_{ij} \p^kE_k + \mu \p^k (\p_k D_{ij} - \p_i D_{kj})\,.\nonumber
\end{eqnarray}
The evolution of the constraint variables becomes
\begin{eqnarray}
\p_t C &=& \p^i C_{ikk} + \mu \p^i\p_i C\,,\label{Eq:evolC1}\\
\p_t C_{ijk} &=& 2\p_{[i} \delta_{j]k} C + 2\mu \p_{[i}
  \p^lC_{|l|j]k}\,.\label{Eq:evolC2}
\end{eqnarray}

By following this procedure, we have lost symmetric hyperbolicity.  It
is essential to check whether the modified systems still admit a
well-posed initial value problem.  We start by showing that both
systems, (\ref{Eq:Et})--(\ref{Eq:Dt}) and
(\ref{Eq:evolC1})--(\ref{Eq:evolC2}), satisfy a necessary condition
for well-posedness, the Petrovskii condition \cite{GKO}.  We will show
that for positive values of the parameter $\mu$ most Fourier
components of the constraint variables are damped.  We then give an
energy estimate for the evolution of the constraints systems and, as a
byproduct, obtain constraint-preserving boundary conditions.  Our
attempts at deriving an estimate for the main system have not been
successful when boundary conditions consistent with the constraints
are imposed.  Nevertheless, we found the one dimensional reduction of
the problem to be well-posed and therefore used it as a tool for
obtaining appropriate boundary conditions for the main system.  The
rationale being that any boundary condition that makes the reduced
problem ill-posed would be unsuitable in higher dimensions.

Let $u_t = P(\p_x)u$ be a system of linear constant coefficient
partial differential equations in $d$ dimensions.  Assume for the
moment that there are no boundaries.  According to the Petrovskii
condition, it must be possible to bound the real parts of the
eigenvalues of the symbol, or Fourier transform, of the differential
operator, $\hat P (i\omega)$, by a real constant $\alpha$, which is
independent of $\omega$.  For scalar equations this condition is also
sufficient.

To compute the symbols of (\ref{Eq:Et})--(\ref{Eq:Dt}) and
(\ref{Eq:evolC1})--(\ref{Eq:evolC2}) we use the substitution $\p_k \to
i\omega_k = i |\omega| \hat\omega_k$, where $\hat\omega_k\hat\omega^k
= 1$.  From the evolution equations we get
\begin{eqnarray*}
\p_t E_{\omega} &=& i |\omega| D_{AA} - \mu
\omega^2 E_{\omega}\,,\\
\p_t E_A &=& - i |\omega| D_{\omega A}\,,\\
\p_t D_{\omega\omega} &=& 0 \,,\\
\p_t D_{\omega A} &=& -i|\omega| E_{A}\,,\\
\p_t D_{A\omega} &=& -\mu\omega^2 D_{A\omega}\,,\\
\p_t D_{AA} &=& 2i|\omega|E_{\omega} - \mu \omega^2 D_{AA}\,,\\
\p_t D_{\hat{AB}} &=& -\mu\omega^2 D_{\hat{AB}}
E_{\omega}\,,
\end{eqnarray*}
where $u_\omega = \hat\omega^k u_k$, $u_A = \hat A^k u_k$, $\hat A_k
\hat B^k = \delta_{AB}$, $\hat\omega_k \hat A^k = 0$, and
$u_{\hat{AB}} = u_{AB} - \frac{1}{2}\delta_{AB}u_{CC}$.  The evolution
of the constraints yields
\begin{eqnarray*}
\p_t C &=& i |\omega| C_{\omega AA} - \mu \omega^2 C\,,\\
\p_t C_{\omega AA} &=& 2i |\omega | C - \mu \omega^2
C_{\omega AA}\,,\\
\p_t C_{\omega \hat{AB}} &=& -\mu \omega^2 C_{\omega \hat{AB}}\,,\\
\p_t C_{\omega A \omega} &=& -\mu \omega^2 C_{\omega A \omega}\,,\\
\p_t C_{AB k} &=& 0\,.
\end{eqnarray*}
The eigenvalues of the Fourier transformed systems form a subset of 
\[
\{\pm i|\omega|, -\mu \omega^2 \pm i\sqrt{2}|\omega|, -\mu \omega^2, 0\}\,.
\]
Hence the Petrovskii condition is satisfied provided that $\mu \ge 0$.
In particular, this shows that for $\mu>0$ the high frequency
components of most constraints are rapidly damped.

We now calculate the time derivative of (\ref{Eq:EC}).  This is given by
\begin{eqnarray}
\frac{d}{dt}E_C &=& \int_{\p\Omega} \left(  C C_{nAA} + \mu C \p_n C
+ \mu C_{nAk}\p_lC^{lAk} \right) \rmd^2 S \nonumber\\
&&- \mu \int_{\Omega} \left(
\p^iC\p_iC + \p_iC^{ijk} \p^lC_{ljk}\right) \rmd^3 x \,.\label{Eq:ECdot}
\end{eqnarray}
Clearly, the volume integral gives a non-positive contribution to the
time derivative of the energy \footnote{Notice that, in the absence of
boundaries, if $C=\rm{const.}$ and $\p^iC_{ijk} = 0$ then the energy
associated with the constraints will remain constant.} and, therefore,
by controlling the boundary term one can ensure that the energy
associated with the constraints is non-increasing.  Boundary
conditions that can be used for this purpose are
\begin{equation}
C = 0, \quad C_{nAk} = 0 \,.\label{Eq:DirC}
\end{equation}
When these 7 conditions are enforced we obtain the inequality
\[
\frac{d}{dt}E_C  = - \mu \int_{\Omega} \left(
\p^iC\p_iC + \p_iC^{ijk} \p^lC_{ljk}\right) \rmd^3 x  \le 0\,.
\]

We now turn to the main system.  As mentioned earlier, in the presence
of boundaries we were not able to obtain an energy estimate using
conditions (\ref{Eq:DirC}).  We therefore look at simple necessary
conditions for well-posedness for the quarter space problem.  We
assume that the boundary is located at $x=0$ and that $\Omega = \{
(x,y,z) \in \mathbb{R}^3 | x \ge 0\}$.  If the variables depend only
on $x$ and $t$, the system becomes
\begin{eqnarray*}
\p_t E_1 &=& +\p_xD_{AA} + \mu \p_x^2 E_1\,,\\
\p_t E_A &=& -\p_xD_{1A}\,,\\
\p_t D_{11} &=& 0\,,\\
\p_t D_{1A} &=& -\p_x E_A \,,\\
\p_t D_{A1} &=& \mu \p_x^2 D_{A1}\,,\\
\p_t D_{AB} &=& \delta_{AB} \p_x E_1 +\mu \p_x^2 D_{AB}\,,
\end{eqnarray*}
where $A = 2, 3$ and $B = 2, 3$.  No matter what boundary conditions
one uses for the main system, they should be such that the one
dimensional reduction of the problem is well-posed.  The equations
above form a mixed hyperbolic-parabolic system.  This can be seen by
identifying $u=\{E_1,D_{Ak},\}$ and $v = \{E_A,D_{1k}\}$, and
comparing with Eq.~(\ref{Eq:mixed}).  The 7 variables belonging to the
parabolic sector require a boundary condition each.
Eq.~(\ref{Eq:DirC}) provides us with the correct number of
Neumann-like boundary conditions.  We are free to specify maximally
dissipative boundary conditions to the 5 variables of the hyperbolic
sector,
\begin{equation}
(w_{A \rm in} - S_{AB} w_{B\rm out})|_{x=0} = g_A\,, \label{Eq:maxdis}
\end{equation}
where $S$ is a sufficiently small $2\times2$ matrix, $g_A$ is
freely specifiable boundary data and
\begin{eqnarray*}
w_{A \rm in} &=& \frac{1}{\sqrt 2}(E_A + D_{1A})\,,\\
w_{A \rm out} &=& \frac{1}{\sqrt 2} (E_A - D_{1A})\,,
\end{eqnarray*}
are the ingoing and outgoing characteristic variables.  This gives a
total of 9 boundary conditions for the main evolution system, 7 of
which ensure constraint preservation
\footnote{The fact that the 7 boundary conditions (\ref{Eq:DirC})
allow for an energy estimate for the constraint propagation system and that any
smaller number of conditions would make this problem ill-posed (for
this purpose it is sufficient to look at its one dimensional
reduction), demonstrates that 7 is the correct number of boundary
conditions for system (\ref{Eq:evolC1})--(\ref{Eq:evolC2}).}.

A description of the discretization of the initial-boundary value
problem and the results of the numerical experiments are given in
Sec.~\ref{Sec:numerics}.

%%%%%%%%%%%%%%%%%%%%%%%%%%%%%%%%%%%%%%%%%%%%%%%%%%%%%%%%%
\section{The KWB system}
\label{Sec:KWB}
%%%%%%%%%%%%%%%%%%%%%%%%%%%%%%%%%%%%%%%%%%%%%%%%%%%%%%%%%

In Ref.~\cite{KWB} Knapp, Walker, and Baumgarte, introduced a
formulation of Maxwell's equations which shares some features with the
Baumgarte-Shapiro-Shibata-Nakamura formulation of Einstein equations
\cite{SN,BS}.  In vacuum and on a Minkowski background in Cartesian
coordinates the equations take the form
\begin{eqnarray}
\p_t A_i &=& -E_i -\p_i \phi \,,\label{Eq:KWB1}\\
\p_t E_i &=& -\p_j\p^jA_i + \p_i \Gamma \,,\label{Eq:KWB2}\\
\p_t \Gamma &=& -\p_j\p^j\phi \,.\label{Eq:KWB3}
\end{eqnarray}
where the fields $A_i$, $E_i$ and $\Gamma$ are subject to the constraints
\begin{eqnarray}
C_E \equiv \p^iE_i = 0\,, \label{Eq:KWB4}\\
C_\Gamma \equiv \p_iA^i - \Gamma = 0 \,.\label{Eq:KWB5}
\end{eqnarray}
The system contains second spatial derivatives and is symmetric
hyperbolic \cite{Gun}.  As the authors of \cite{KWB} pointed out, in
this system the constraints propagate according to the wave equation
\begin{eqnarray}
\p_t C_E &=& -\p_j\p^j C_\Gamma \,,\label{Eq:C_E}\\
\p_t C_\Gamma &=& -C_E \,,\label{Eq:C_G}
\end{eqnarray}
which is to be contrasted with the static constraint evolution of the
original Maxwell system, Eqs.~(\ref{Eq:max1})--(\ref{Eq:max4}).

The constraint system (\ref{Eq:C_E})--(\ref{Eq:C_G}) conserves the
following energy
\begin{equation}
E_C = \frac{1}{2}\int_\Omega (C_E^2 + \p^iC_\Gamma \p_iC_\Gamma )
\rmd^3 x \,.\label{Eq:E_C}
\end{equation}
If one sets $\phi = 0$ and modifies the evolution equations according
to
\begin{eqnarray}
\p_t A_i &=& \ldots + \mu \p_i C_\Gamma\\
&=& -E_i + \mu \p_i\p_j A^j - \mu \p_i \Gamma\,,\nonumber\\
\p_t E_i &=& \ldots + \mu \p_i C_E\\
 &=& -\p_j\p^jA_i + \p_i \Gamma + \mu \p_i\p^jE_j\,,\nonumber\\
\p_t \Gamma &=& \lambda C_\Gamma = \lambda \p_iA^i - \lambda
\Gamma\,. \label{Eq:Gamma_dot}
\end{eqnarray}
which can be identified with Eqs.~(22)--(23) of Fiske \cite{F},
the evolution of the constraints becomes
\begin{eqnarray}
\p_t C_E &=& -\p^i\p_i C_\Gamma + \mu \p^i\p_i C_E\,, \label{Eq:dCE}\\
\p_t C_\Gamma &=& -C_E +\mu \p^i\p_i C_\Gamma -\lambda C_\Gamma\,.
\label{Eq:dCG}
\end{eqnarray}
The evolution equations were modified so that (\ref{Eq:dCE}) and
(\ref{Eq:dCG}) would acquire damping and friction terms.

Rather than estimating the energy (\ref{Eq:E_C}) we observe that, if
$\mu>1/2$, the constraint propagation system is strongly parabolic and
an energy estimate for the constraint system can be obtained directly
in $L_2$.  The time derivative of
\begin{equation}
E_C = \frac{1}{2} \int_\Omega \left( C_E^2 + C_\Gamma^2 \right) \rmd^3
x
\end{equation}
is
\begin{eqnarray}
&& \frac{d}{dt}E_C = \int_{\p\Omega} \left( - C_E \p_n C_\Gamma + \mu
C_E\p_n C_E + C_\Gamma \p_n C_\Gamma \right) \rmd^2S\nonumber\\
&&+ \int_\Omega \left( \p^iC_E \p_iC_\Gamma - \mu \p^i C_E \p_iC_E -
\mu\p^iC_\Gamma \p_i C_\Gamma \right) \rmd^3 x \nonumber\\
&&+ \int_\Omega \left( - C_\Gamma C_E - \lambda C_\Gamma^2 \right)
\rmd^3 x\,. \label{Eq:volterm}
\end{eqnarray}
The use of Dirichlet boundary conditions, 
\begin{equation}
C_E = C_\Gamma = 0\,, \label{Eq:DirC2}
\end{equation}
ensures that the boundary term of (\ref{Eq:volterm}) vanishes.
Furthermore, the volume term containing spatial derivatives is
non-positive.  Thus, an energy estimate can be obtained
\footnote{To ensure that the volume term of Eq.~(\ref{Eq:volterm})
containing undifferentiated constraints is semidefinite negative, so
that the inequality $\frac{d}{dt} E_C \le 0$ holds, one could add the
term $-C_E$ to the right hand side of Eq.~(\ref{Eq:Gamma_dot}).}.

We now place a boundary at $x=0$ so that $\Omega =
\{(x,y,z)\in\mathbb{R}^3| x \ge 0\}$.  By ignoring any $y$ and $z$
dependence the hyperbolic-parabolic character of the system becomes
evident (the introduction of the auxiliary variable $W_B = \p_x A_B$
may be helpful for this analysis)
\begin{eqnarray*}
\p_t A_1 &=& -E_1 + \mu \p_x^2 A_1 - \mu \p_x \Gamma\,,\\
\p_t A_B &=& -E_B\,,\\
\p_t E_1 &=& -\p_x^2 A_1 + \p_x \Gamma + \mu \p_x^2 E_1\,,\\
\p_t E_B &=& -\p_x^2 A_B\,,\\
\p_t \Gamma &=& \lambda \p_xA_1 -\lambda \Gamma\,.
\end{eqnarray*}
For $\mu>1/2$ we can identify a parabolic sector $\{A_1,E_1\}$ and a
hyperbolic sector $\{A_B,E_B,\Gamma\}$.  Dirichlet boundary conditions
on the constraints, $C_E = 0$, $C_\Gamma = 0$, correspond to
Neumann-like boundary conditions on the main variables of the
parabolic sector.  Maximally dissipative boundary conditions can be
imposed on the variables of the hyperbolic sector,
\begin{equation}
(\p_xA_B+E_B)|_{x=0} = 0\,. \label{Eq:kwbsomm}
\end{equation}

Numerical experiments with the modified KWB system indicate that the
discrete approximation of the initial-boundary value problem is
stable.  More details are given in the next Section.  In Appendix
\ref{App:discretization} we analyze the accuracy and stability of
discretized maximally dissipative boundary conditions, such as
Eq.~(\ref{Eq:kwbsomm}), for the one dimensional wave equation in
second order form.

%%%%%%%%%%%%%%%%%%%%%%%%%%%%%%%%%%%%%%%%%%%%%%%%%%%%%%%%%%%%
\section{Numerical implementation}
\label{Sec:numerics}
%%%%%%%%%%%%%%%%%%%%%%%%%%%%%%%%%%%%%%%%%%%%%%%%%%%%%%%%%%%%

To discretize our problems we use the method of lines, whereby space
is discretized while time is left continuous.  We then integrate the
resulting system of ordinary differential equations using 4th order
Runge-Kutta.  We take our domain to be $\Omega = [0,1]^3$ and use
periodic boundary conditions in $y$ and assume no dependence on $z$.
We introduce a grid such that the points $(0,j,k)$ belong to the
boundary, $x=0$.  In the interior we approximate the first spatial
derivatives with centered differencing operators and the second
spatial derivatives according to
\[
\p_{x^l}\p_{x^m} u \to \left\{ \begin{array}{cc} 
D_{+x^l}D_{-x^l} u_{ijk} & l=m\\
D_{0x^l}D_{0x^m} u_{ijk} & l\neq m
\end{array}
\right.
\]
where $hD_+ u_j = u_{j+1}-u_j$, $hD_-u_j = u_j - u_{j-1}$, and $2D_0 =
D_+ + D_-$.  Thus, the discretization in the interior is second order
accurate.

The discretization of the boundary conditions is based on algorithms
for which convergence proofs are available, at least for simple toy
model problems \cite{GKO}.  For example, one can show that for the
one-way advective equation, $u_t = u_x$, $x \ge 0$, one can use
one-sided differencing at the boundary, $\frac{du_0}{dt} = D_+u_0$.
For the heat equation, $u_t = u_{xx}$, $x\ge 0$, with Neumann boundary
condition $u_x(t,0) = 0$, one can use the approximation $D_0u_0(t) =
0$ which, when combined with the evolution equation at $i = 0$ becomes
$\frac{du_0}{dt} = \frac{2}{h} D_+ u_0$.  Both cases give global
second order convergence.

To simplify the numerical implementation of the boundary conditions we
found it convenient to introduce a ghost zone consisting of an extra
grid point, $i=-1$, to the left of the boundary, so that derivatives
could be computed at $i=0$ as if it were an interior point.  In the
fat Maxwell case we use the following prescription to populate the
ghost zone: we use the Neumann-like boundary conditions
(\ref{Eq:DirC}) and use second order extrapolation for the remaining
fields, $\{E_A,D_{1k}\}$. Explicitly,
\begin{eqnarray*}
0 &=& D_{0x}E_{1;0jk} + D_{0y}E_{2;0jk} + D_{0z}E_{3;0jk}\,,\\
0 &=& D_{0x}D_{Am;0jk} - D_{0A}D_{1m;0jk}\,,\\
0 &=& D_{+x}D_{-x}E_{A;0jk}\,,\\
0 &=& D_{+x}D_{-x}D_{1m;0jk}\,.
\end{eqnarray*}
Note that when combined with a second order centered difference at
$i=0$, second order extrapolation is equivalent to first order
one-sided differencing.  Eq.~(\ref{Eq:maxdis}) with $S_{AB} =
g_A = 0$ is then used to prescribe the incoming fields of the hyperbolic part.
The new variables are computed by solving the system
\begin{eqnarray*}
E_A^{(\rm new)} + D_{1A}^{(\rm new)} &=& 0\\
E_A^{(\rm new)} - D_{1A}^{(\rm new)} &=& E_A^{(\rm old)} - D_{1A}^{(\rm old)} 
\end{eqnarray*}

In the KWB case the boundary treatment is slightly different due to
the hyperbolic role that some second spatial derivatives play in the
evolution equations.  The theory of difference approximations for
second order hyperbolic systems is much less developed than the
correspondent one for first order systems (for some recent developments
see Ref.~\cite{KPY}).  Based on the energy method for semi-discrete
systems, in Appendix \ref{App:discretization} we give a proof of
second order convergence for a particular discretization of the one
dimensional wave equation, $u_{tt} = u_{xx}$, $x \ge 0$, with
Sommerfeld boundary conditions, $u_t(t,0) = u_x(t,0)$.  This result is
used in the discretization of boundary condition (\ref{Eq:kwbsomm}).

As in the fat Maxwell case, we introduce a ghost zone and populate it
using the following equations
\begin{eqnarray*}
0 &=& D_{0x}A_{1;0,jk} + D_{0y}A_{2;0jk} + D_{0z}A_{3;0jk} - \Gamma_{0jk}\,,\\
0 &=& D_{0x}A_{B;0jk} + E_{B;0,jk}\,,\\
0 &=& D_{0x}E_{1;0,jk} + D_{0y}E_{2;0jk} + D_{0z}E_{3;0jk}\,,\\
0 &=& D_{+x}D_{-x}E_{B;0jk}\,,\\
0 &=& D_{+x}D_{-x}\Gamma_{0jk}\,.
\end{eqnarray*}
The first and third conditions correspond to the vanishing of the
constraints, Eq.~(\ref{Eq:DirC2}).  The second condition is the
discretization of Eq.~(\ref{Eq:kwbsomm}), done according to the
prescription (\ref{Eq:dphi0}).  The last two conditions
correspond to second order extrapolation.

\begin{figure}[t]
\begin{center}
\includegraphics*[height=6cm]{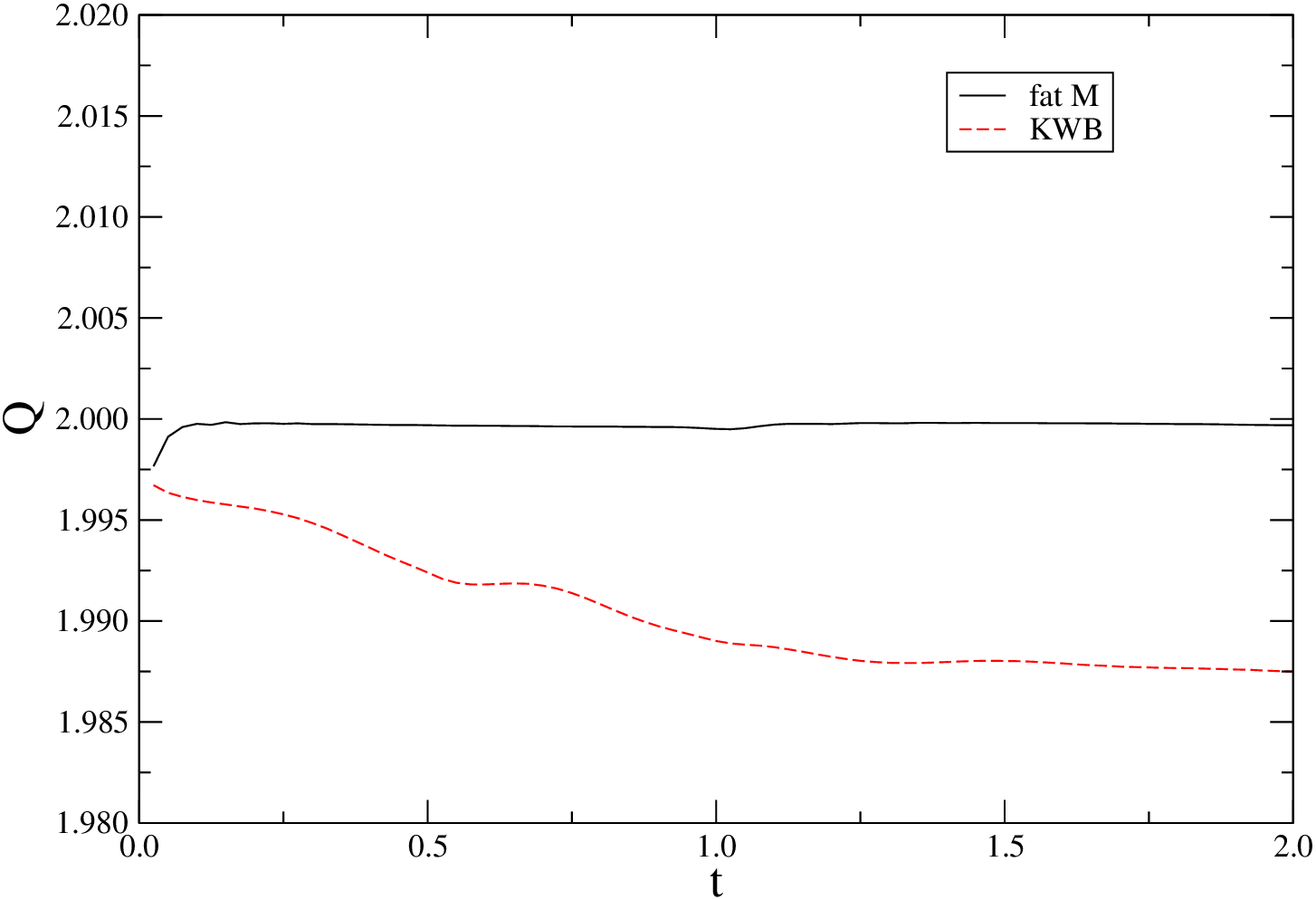}
\caption{To test the stability and accuracy of our discretization we
  perform a grid refinement test.  The domain is $\Omega = [0,1]^2$
  with no dependence in the $z$ direction and periodic boundary
  conditions in the $y$ direction.  At $x=0$ and $x=1$ we use the
  discrete boundary conditions described in Sec.~\ref{Sec:numerics}.
  For both the fat Maxwell and KWB systems we plot $Q = \log_2 (\|
  u_h-u_e\| / \| u_{h/2} - u_e \|)$ where $u_h$ and $u_{h/2}$
  represent the numerical solutions computed with a mesh spacing of
  $h=1/100$ and $u_e$ the exact solution, which is given by a wave
  travelling in an oblique direction with respect to the boundary.
  For the fat Maxwell system it is obtained from the potentials $\phi
  = 0$ and $\vec A = \frac{1}{\pi} \sin [\pi (x+2y+\sqrt{5}t)]
  (0,0,1)$, while for the KWB system from $\phi = 0$ and $\vec A =
  \frac{1}{\pi} \sin [\pi (x+2y+\sqrt{5}t)] (2,-1,0)$. In the fat
  Maxwell case the time step, $k$, is chosen so that $4\mu k/h^2 = 1$,
  where $h$ is the spatial mesh spacing, and we set $\mu = 1/10$.  In
  the KWB case we used $2\mu k/h^2 = 1$, $\mu = 1$ and $\lambda = 10$.
  Values of $Q$ close to 2 indicate second order convergence for the
  main variables.}
\label{Fig:gridref}
\end{center}
\end{figure}

\begin{figure}[t]
\begin{center}
\includegraphics*[height=6cm]{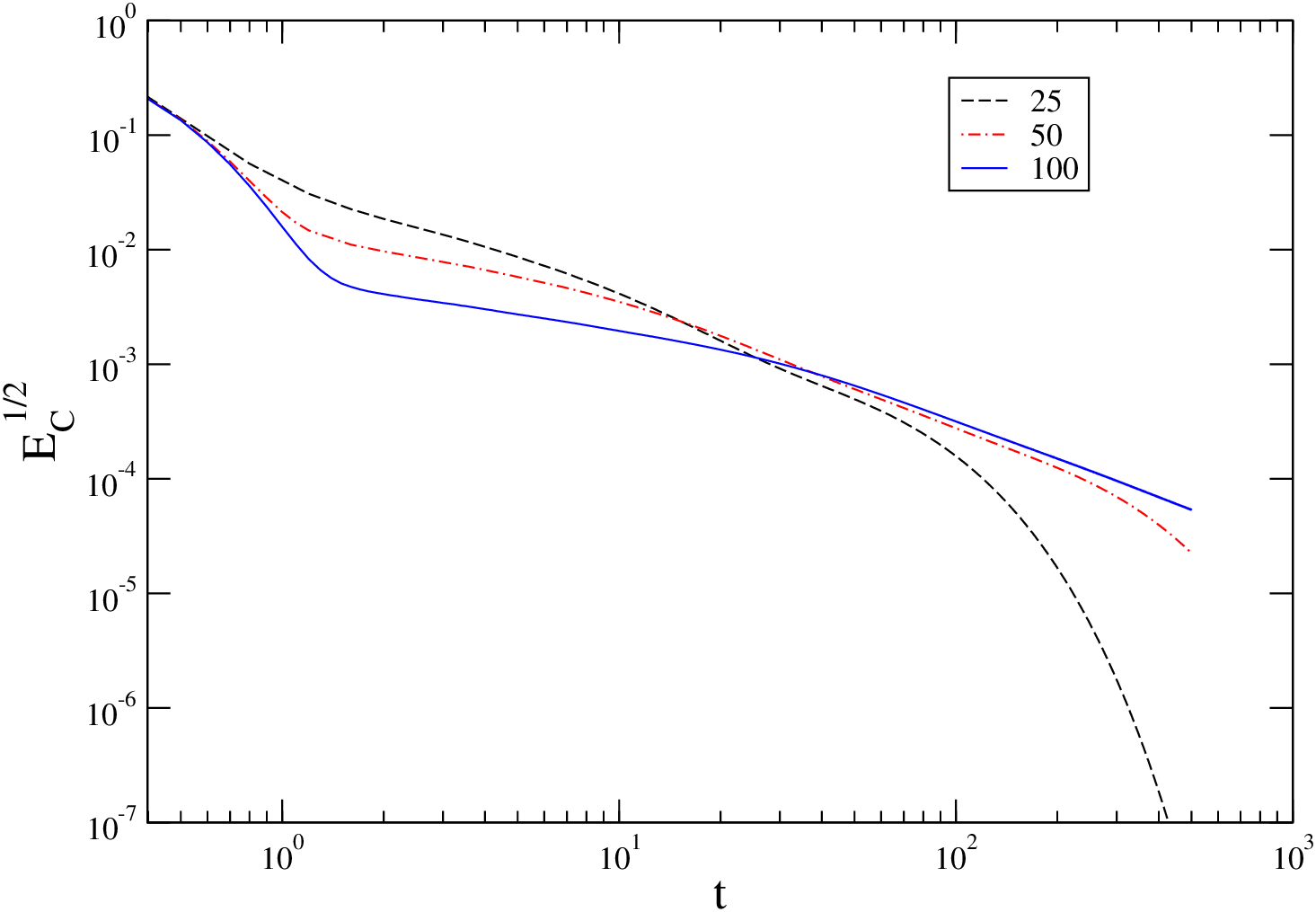}
\includegraphics*[height=6cm]{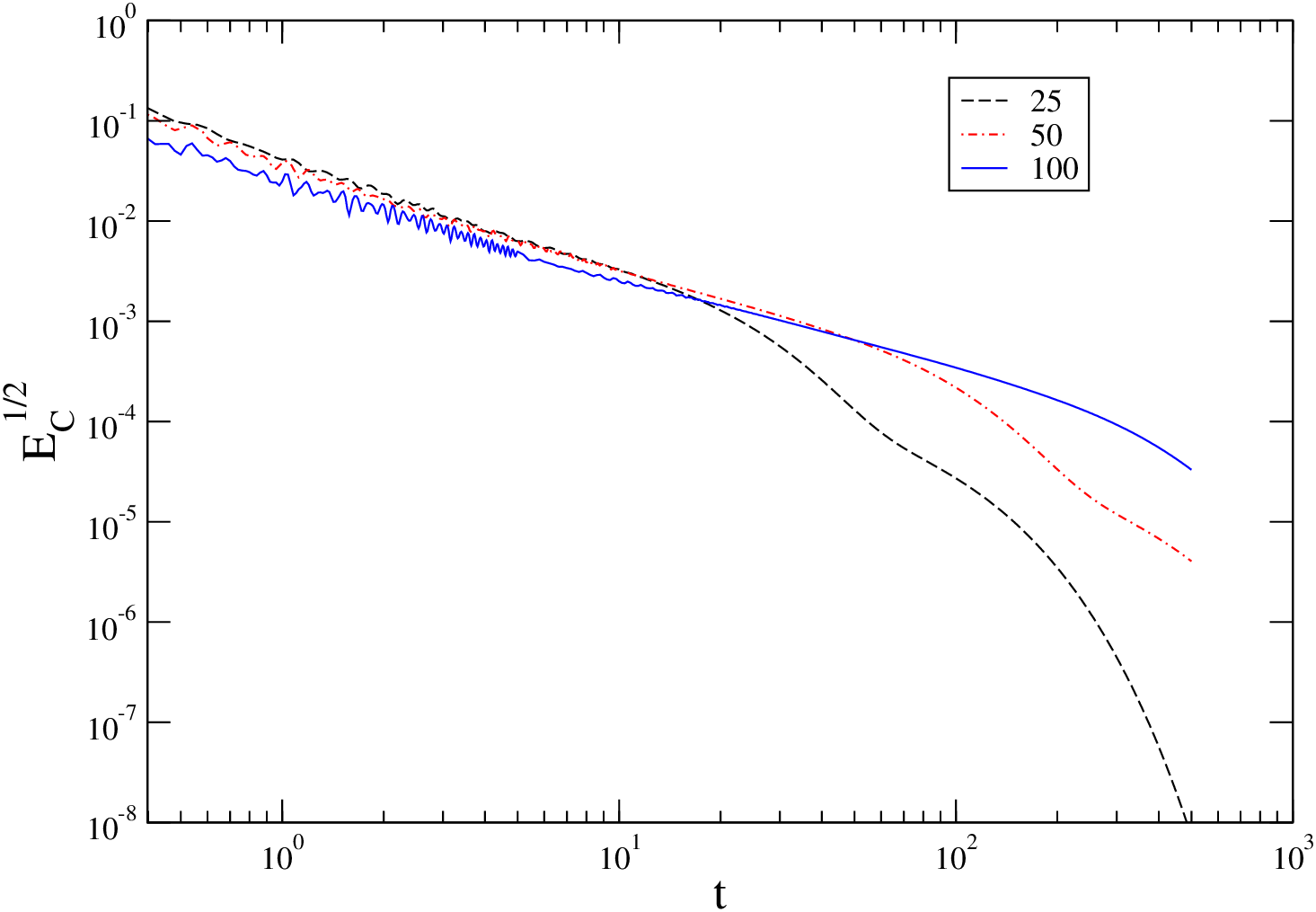}
\caption{We plot the value of $E_C^{1/2}$, Eqs.~(\ref{Eq:EC}) and
  (\ref{Eq:E_C}), for the fat Maxwell system (top) and the KWB system
  (bottom) for three different resolutions.  The curves are labeled
  according to the number of grid points used in each spatial
  direction.  We give constraint violating initial data consisting of
  a sufficiently smooth pulse of compact support to all fields.  In
  both cases the domain is $\Omega = [0,1]^2$, having dropped the
  dependence in the $z$ direction, and we impose periodic boundary
  conditions in the $y$ direction.  At $x=0$ and $x=1$ we use the
  boundary conditions outlined in Sec.~\ref{Sec:numerics}.  The values
  of the parameters and the size of the time step $k$ are chosen as in
  Fig.~\ref{Fig:gridref}.  The noise in the curves of the KWB case are
  likely due to the fact that the energy being monitored,
  Eq.~(\ref{Eq:E_C}), contains spatial derivatives which are
  discretized using $D_0$ operators.}
\label{Fig:decay}
\end{center}
\end{figure}

\begin{figure}[t]
\begin{center}
\includegraphics*[height=6cm]{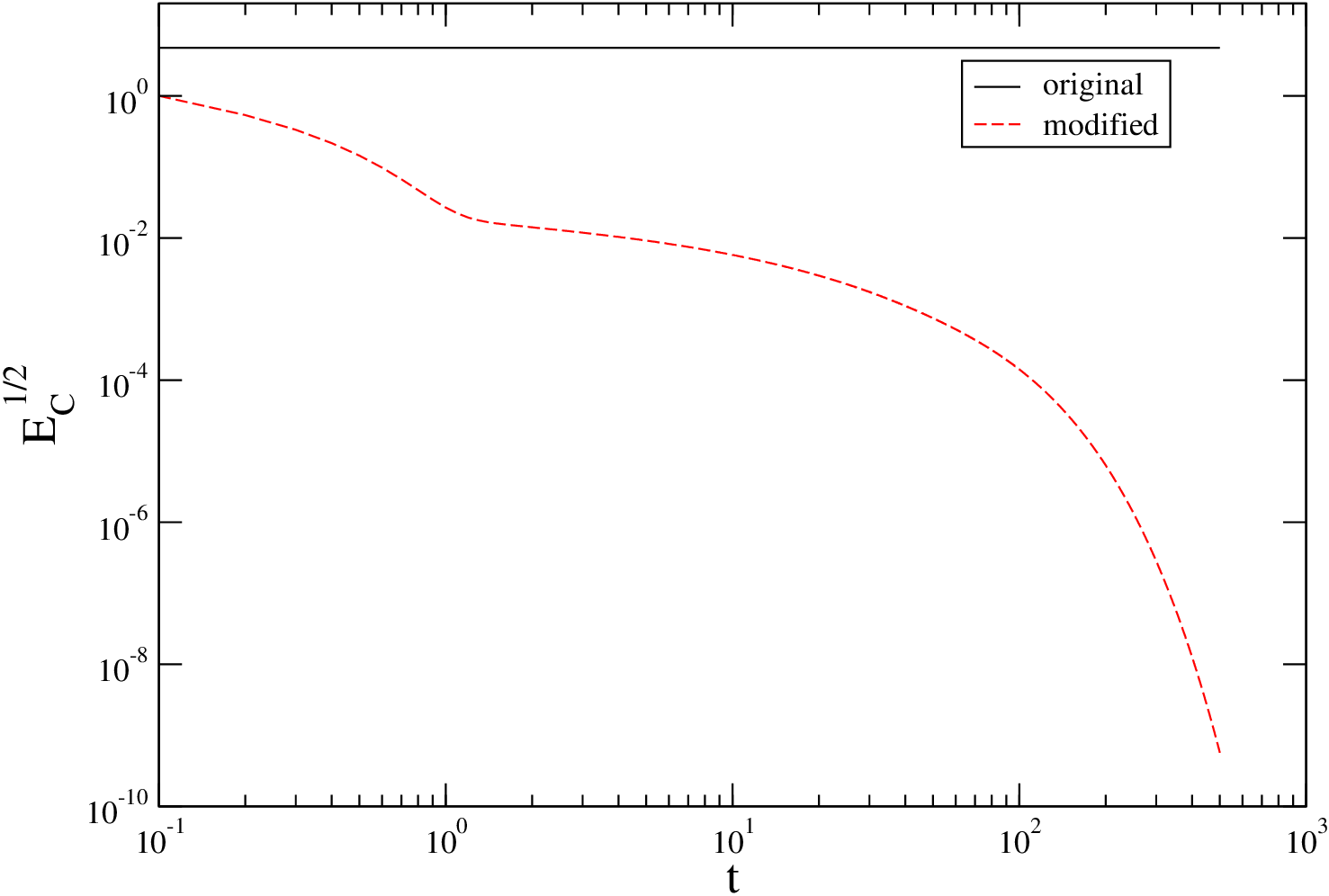}
\includegraphics*[height=6cm]{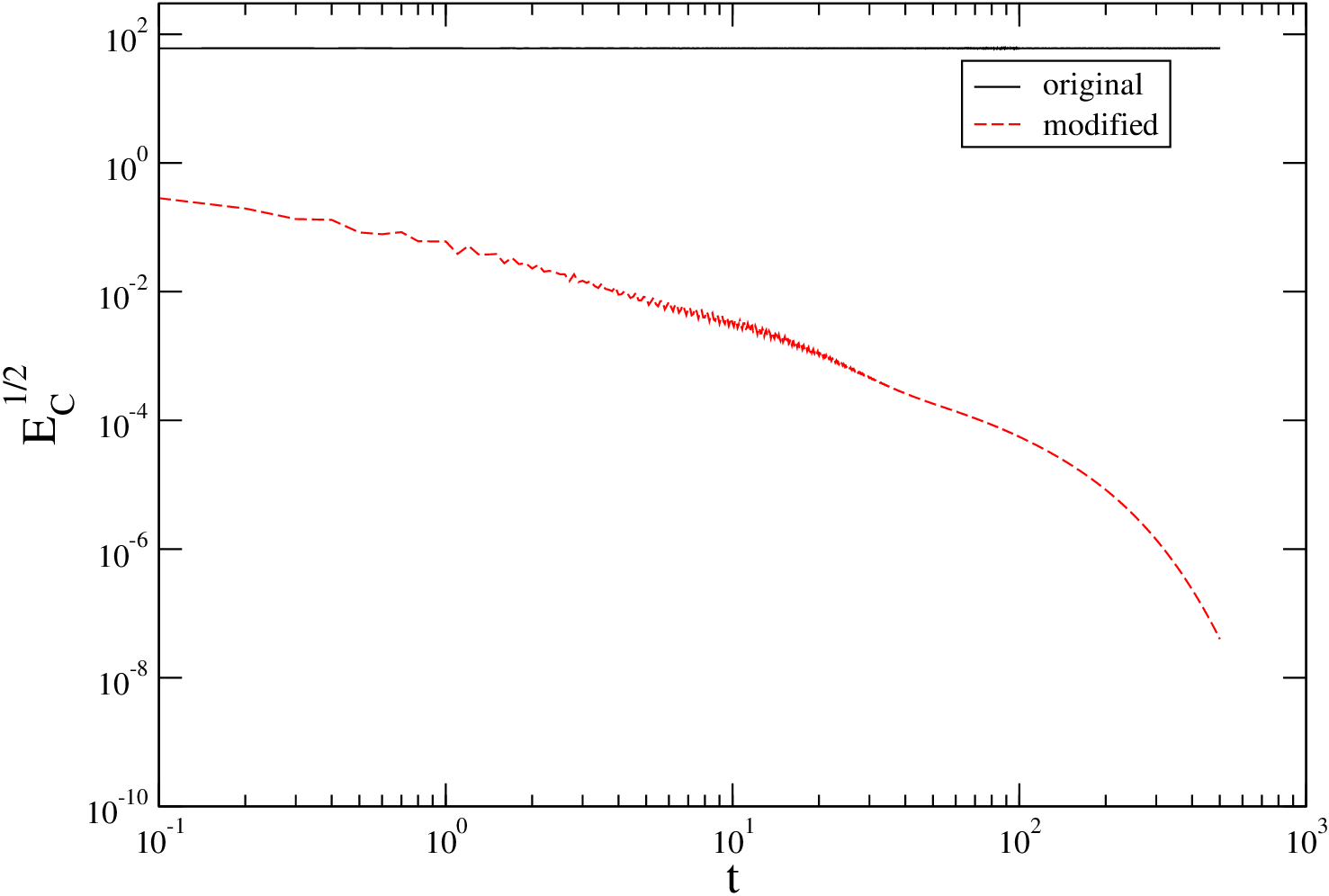}
\caption{At a fixed resolution of $h= 1/50$ and with periodic boundary
  conditions and no dependence in $z$ we compare the decay of the
  constraint energies, Eqs.~(\ref{Eq:EC}) and (\ref{Eq:E_C}), between
  the modified and unmodified systems for the fat Maxwell (top) and
  the KWB (bottom) systems.  As in Fig.~\ref{Fig:decay} the initial
  data is a constraint violating pulse of compact support to all
  fields.  Although this cannot be seen from the plots, at time $t=0$
  the constraint energy of the modified and unmodified systems
  coincide.}
\label{Fig:comparison}
\end{center}
\end{figure}

We treat the $x=1$ boundary in a similar way.  Having verified that
our discretizations are second order convergent for both the fat
Maxwell and the KWB systems, see Fig.~\ref{Fig:gridref}, we test our
schemes with constraint violating initial data corresponding to a
sufficiently smooth pulse of compact support to all fields and perform
runs at three different resolutions. Given that in the fat Maxwell
system there is no restriction on the value of $\mu$, we choose
something reasonably small, $1/10$, so that we can take larger time
steps.  In the KWB case, instead, we must choose $\mu > 1/2$ otherwise
it would not be clear that one could use $C_E = C_\Gamma = 0$ as
boundary conditions for the constraints.  We opt for $\mu = 1$.  The
other parameter that is present in the KWB formulation is $\lambda$.
This is subject to no restriction, does not affect the choice of time
step (for sufficiently high resolutions), and, if positive, contributes to the
damping.  We choose $\lambda = 10$.  Not only do we
observe decay of the constraints, as shown in Fig.~\ref{Fig:decay},
but we also see no signs of instability in the main system.  Finally,
in Fig.~\ref{Fig:comparison}, we compare the constraint decay of the
modified systems with that of the original ones.

%%%%%%%%%%%%%%%%%%%%%%%%%%%%%%%%%%%%%%%%%%%%%%%%%%%%%%%%%
\section{Concluding remarks}
\label{Sec:conclusion}
%%%%%%%%%%%%%%%%%%%%%%%%%%%%%%%%%%%%%%%%%%%%%%%%%%%%%%%%%

In this work we investigated the effect of adding spatial derivatives
of constraints to the main evolution system.  We studied the
initial-boundary value problem for two different formulations of
Maxwell's equations, the fat Maxwell system and the
Knapp-Walker-Baumgarte formulation.  We noticed that the modification
alters the character of the systems, transforming them into mixed
hyperbolic-parabolic problems.  Parabolicity introduces two remarkable
effects.  First, it damps the high frequency components of constraint
violations (whereas in the KWB case the damping affects all
constraints, in the fat Maxwell system some constraints are not
damped).  Second, it allows for Dirichlet boundary conditions for the
constraints which naturally translate into Neumann-like boundary
conditions for the main system.  These boundary conditions are
constraint-preserving and, if the evolution of the constraints can be
made strongly parabolic, their expression is simpler than that of the
original hyperbolic system.  Furthermore, whereas other techniques,
such as the $\lambda$-system \cite{lambda}, require a substantial
enlargement of the main system, this method leaves the number of
variables unchanged.  Although we were unsuccessful in obtaining an
energy estimate for the main evolution systems, our numerical tests
not only confirmed that the constraints are damped, but also indicated
that the discretized systems are stable.

The stringent requirement on the time step $k$ is possibly the
greatest drawback of this technique.  Whereas for hyperbolic problems
$k$ typically has to be proportional to the mesh spacing, $h$, in
parabolic or mixed hyperbolic-parabolic problems it has to be
proportional to $h^2$.  To circumvent this restriction one could use
implicit or semi-implicit schemes.  However, we have not yet
experimented with such schemes.

The damping terms introduced by the spatial derivatives of constraints
vanish on physical (constraint satisfying) solutions.  This indicates
that dissipation will not detrimentally affect such solutions and,
although constraint violations can travel at an infinite speed, any
constraint satisfying solution will propagate information at a finite
speed.  Of course, at the numerical level, the numerical speed of
propagation usually only coincides with the continuum one up to
truncation errors and it is even possible for high frequency modes to
travel in the wrong direction \cite{KO}.

It is important to realize that in the nonlinear or variable
coefficient case, extra care is required for the successful
application of this technique.  The coefficients in front of the
parabolic terms should never become negative, as this would
immediately allow for unbounded growth and lead to instabilities.
Artificial dissipation of the Kreiss-Oliger type should still be added
to the right hand side in order to damp high frequency components of
the hyperbolic sector.  In addition, this technique, just like most of
the available cures for constraint growth, cannot be expected to cure
instabilities, such as ``gauge instabilities'', that are not
constraint violating.

The technique illustrated in this work requires a different treatment
of the excision boundaries, often used in numerical simulations of
spacetimes containing black holes.  Whereas in hyperbolic formulations
no boundary conditions are needed at the excision surface, provided
that this is purely outflow, parabolicity demands boundary conditions.
Our expectation is that the use of Dirichlet boundary conditions for
the constraints at the excision boundary will help to prevent the
growth of constraints near this troublesome region.

In this work we stressed how parabolicity can be obtained by adding
spatial derivatives of constraints to the evolution equations.  For
systems, such as the ones studied in this paper, that have first order
constraints, this is the only way it can be achieved.  However, for
systems containing constraints with second spatial derivatives (like
the Hamiltonian constraint of General Relativity) one may obtain
parabolicity by appropriately adding such constraints without further
differentiation.  This approach of adding constraints and/or spatial
derivatives of constraints to the evolution equations could be pushed
further and one could consider the possibility of adding higher order
spatial derivatives of constraints.  However, in general, adding
second spatial derivatives of first order constraints or first spatial
derivatives of second order constraints will not damp high frequency
components.  It will only render the constraint propagation more
dispersive.  On the other hand, adding third spatial derivatives of
first order constraints or second spatial derivatives of second order
constraints can help to dissipate the high frequencies.  Certainly, it
is preferable to avoid increasing the order of the formulation, as
this would complicate the treatment of boundaries.

Although the problems to which our technique was applied do not
present the full range of difficulties contained in Einstein's
equations, based on the success of the tests that we presented we hope
that the technique may prove helpful in a more complex situation.  Its
effectiveness in non-linear problems of General Relativity will be
assessed in future works.

%%%%%%%%%%%%%%%%%%%%%%%%%%%%%%%%%%%%%%%%%%%%%%%%%%%%%%%%%
% Acknowledgments
%%%%%%%%%%%%%%%%%%%%%%%%%%%%%%%%%%%%%%%%%%%%%%%%%%%%%%%%%

\begin{acknowledgments}
We wish to thank David Fiske, Carsten Gundlach, Luis Lehner, David
Neilsen, and Olivier Sarbach for helpful discussions and suggestions.
This research was supported by a Marie Curie Intra-European Fellowship
within the 6th European Community Framework Program.
\end{acknowledgments}

%%%%%%%%%%%%%%%%%%%%%%%%%%%%%%%%%%%%%%%%%%%%%%%%%%%%%%%%%
\appendix
%%%%%%%%%%%%%%%%%%%%%%%%%%%%%%%%%%%%%%%%%%%%%%%%%%%%%%%%%

%%%%%%%%%%%%%%%%%%%%%%%%%%%%%%%%%%%%%%%%%%%%%%%%%%%%%%%%%
\section{A discretization of the second order wave equation with Sommerfeld
  boundary conditions}
\label{App:discretization}
%%%%%%%%%%%%%%%%%%%%%%%%%%%%%%%%%%%%%%%%%%%%%%%%%%%%%%%%%

Consider the wave equation, 
\begin{equation}
\p^2_t \phi (t,x) = \p^2_x \phi (t,x) \,,\label{Eq:waveexa1}
\end{equation}
on the real
line with appropriate initial data,
\begin{eqnarray}
\phi(0,x) &=& f^{(1)}(x)\,,\label{Eq:waveexa2}\\
\p_t \phi(0,x) &=& f^{(2)}(x)\,. \label{Eq:waveexa3}
\end{eqnarray}  
The analysis that follows can be readily applied to the first order in
time and second order in space formulation
\begin{eqnarray*}
\p_t \phi &=& T \,,\\
\p_t T &=& \p_x^2 \phi \,.
\end{eqnarray*}
As an immediate consequence of integration by parts we have that the
energy $E = \int ((\p_t\phi)^2 + (\p_x\phi)^2)\rmd x$ is conserved by
any solution of the wave equation.  If we discretize the second space
derivative with the second order accurate centered differencing
operator, obtaining the semi-discrete system
\begin{equation}
\frac{d^2\phi_j}{dt^2} = D_+D_- \phi_j\,, \label{Eq:wave_semi}
\end{equation}
and observe that the one-sided finite difference operators
$D_{\pm}$ satisfy the summation by parts rule
\[
(u,D_\pm v)_{-\infty,+\infty} = - (D_\mp u, v)_{-\infty,+\infty}\,,
\]
where
\[
(u,v)_{r,s} = \sum_{j=r}^s  u_j v_j h\,,
\]
it follows that any solution of (\ref{Eq:wave_semi}) conserves the discrete 
energy
\begin{equation}
E = \left(\frac{d\phi}{dt},\frac{d\phi}{dt}\right)_{-\infty,+\infty} + (D_+\phi,D_+\phi)_{-\infty,+\infty}\,.
\end{equation}
Note that the similar expression
\[
E_0 = \left(\frac{d\phi}{dt},\frac{d\phi}{dt}\right)_{-\infty,+\infty} + (D_0\phi,D_0\phi)_{-\infty,+\infty}
\]
which uses centered instead of one-sided differencing is not a
conserved quantity, unless the right hand side of
Eq.~(\ref{Eq:wave_semi}) is replaced by $D_0^2 \phi_j$.  However, this
would require a larger stencil and would not dissipate the highest
frequency mode, $\phi_j = (-1)^j$.

We now introduce an artificial boundary at $x=0$ and take our domain
to be $x \ge 0$.  We are interested in finding approximations for the
homogeneous Sommerfeld boundary condition
\begin{equation}
\phi_t(t,0) = \phi_x(t,0) \,,\label{Eq:Sommerfeld}
\end{equation}
which lead to a second order accurate scheme. 
The initial data are assumed to be compatible with the boundary condition,
i.e., 
\[
\frac{\p^n}{\p x^n} \left.\left(f^{(2)}(x) - \p_x f^{(1)}(x)\right)\right|_{x=0} = 0\,.
\]
Consider the approximation 
\begin{equation}
\frac{d\phi_0}{dt} = D_0\phi_0\,. \label{Eq:dphi0}
\end{equation}
This requires the introduction of $\phi_{-1}$.  However, by combining 
(\ref{Eq:dphi0}) with the evolution equation at $j=0$,
\[
\frac{d^2 \phi_0}{dt^2}= D_+D_- \phi_0\,,
\]
we can eliminate $\phi_{-1}$ from the semi-discrete system.  We obtain
\begin{eqnarray}
\frac{d^2 \phi_j}{dt^2} &=& D_+D_- \phi_j\,,\qquad j = 1, 2, 3, \ldots\,,\label{Eq:wave1}\\
\frac{d^2 \phi_0}{dt^2} &=& 2\left(D_+\phi_0 - \frac{d\phi_0}{dt}\right)/h\,, \label{Eq:bcondord2}
\end{eqnarray}
with initial data
\begin{eqnarray}
\phi_j(0) &=& f^{(1)}(x_j)\,,\label{Eq:id1}\\
\frac{d\phi_j}{dt}(0) &=& f^{(2)}(x_j)\,.\label{Eq:id2}
\end{eqnarray}

We conclude this appendix by proving that the semi-discrete
approximation (\ref{Eq:wave1})--(\ref{Eq:id2}) of the initial-boundary
value problem (\ref{Eq:waveexa1})--(\ref{Eq:waveexa3}) and
(\ref{Eq:Sommerfeld}) is stable and second order convergent.

We denote by $\phi_e(t,x)$ the solution of the continuum problem
(\ref{Eq:waveexa1})--(\ref{Eq:waveexa3}) and (\ref{Eq:Sommerfeld}) and
$\phi_j(t)$ the grid function satisfying the semi-discrete
approximation (\ref{Eq:wave1})--(\ref{Eq:id2}).  The norm with respect
to which we wish to prove stability and convergence is the one
associated with the scalar product
\begin{equation}
2 (u,v)_h  = \frac{1}{2} \frac{du_0}{dt}\frac{dv_0}{dt} h +
\left(\frac{du}{dt},\frac{dv}{dt}\right)_{1,+\infty} +
(D_+u,D_+v)_{0,+\infty}\,. \label{Eq:sp}
\end{equation}
The fact that the time derivative of
\[
2E = \frac{1}{2} \frac{d\phi_0}{dt}^2 h +
\left(\frac{d\phi}{dt},\frac{d\phi}{dt}\right)_{1,+\infty} +
(D_+\phi,D_+\phi)_{0,+\infty}
\]
is
\[
\frac{d}{dt}E = -\left(\frac{d\phi_0}{dt}\right)^2 \le 0
\]
proves that the the approximation is stable.  To prove convergence we
introduce the error grid-function, $w_j(t) = \phi_j(t) -
\phi_{e}(t,x_j)$, and observe that it satisfies
\begin{eqnarray*}
\frac{d^2 w_j}{dt^2}&=& D_+D_- w_j + F_j \,,\qquad j = 1,2,\ldots\,,\\
\frac{d^2 w_0}{dt^2}&=& 2\left(D_+ w_0 - \frac{dw_0}{dt} \right) /h +
F_0\,,\\
w_j(0) &=& 0\,,\qquad j = 0, 1, 2, \ldots\,,\\
\frac{dw_j}{dt}(0) &=& 0\,,\qquad j = 0, 1, 2, \ldots\,,
\end{eqnarray*}
where $F_j = {\cal O}(h^2)$ for $j=1,2,\ldots$ and $F_0 = {\cal
O}(h)$.  Provided that higher derivatives of the exact solution are
appropriately bounded, we can estimate the norm of the error,
\begin{eqnarray*}
\frac{d}{dt}\|w\|^2_h &=& \frac{1}{2} \frac{dw_0}{dt}\frac{d^2w_0}{dt^2}h
+ \left( \frac{dw}{dt},\frac{d^2w}{dt^2}\right)_{1,+\infty} \\
&&+\left(D_+w,D_+\frac{dw}{dt} \right)_{0,+\infty} \\
&=& -\left(\frac{dw_0}{dt}\right)^2 + \frac{1}{2} \frac{dw_0}{dt}F_0h
+ \left(\frac{dw}{dt},F\right)_{1,+\infty} \\
&\le& F_0^2 h^2 + \|w\|^2_h + \|F\|^2_{1,+\infty}\,.
\end{eqnarray*}
Integrating, we obtain the inequality
\begin{eqnarray*}
\|w(t)\|^2_h &\le& \int_0^t e^{t-\tau} F_0^2(\tau)h^2 \rmd\tau +
\int_0^te^{t-\tau} \|F(\tau)\|^2_{1,+\infty}\rmd\tau \\
&=& {\cal O}(h^4)
\end{eqnarray*}
showing that the scheme is second order convergent with respect to the
norm induced by the scalar product (\ref{Eq:sp}).

It is interesting to note that the simpler approximation
\begin{eqnarray}
\frac{d^2 \phi_j}{dt^2} &=& D_+D_-\phi_j \,, \qquad j = 1, 2, 3, \ldots\,,\label{Eq:wave2}\\
\frac{d \phi_0}{dt} &=& D_+ \phi_0 \,,\label{Eq:bcond1}
\end{eqnarray}
although stable, reduces the global accuracy of the scheme to first
order. A similar calculation to the one in the proof leads to the
inequality
\[
\|w(t)\|_h^2 \le {\cal O}(h^2)\,.
\]

%%%%%%%%%%%%%%%%%%%%%%%%%%%%%%%%%%%%%%%%%%%%%%%%

\end{document}